\documentclass[preprint]{revtex4}
\usepackage{amsmath,amsfonts,amssymb}
\usepackage[dvips]{graphics}

\begin{document}
\title{Prediction and verification of indirect interactions in densely interconnected regulatory networks}
\author{Koon-Kiu Yan$^{1,2}$,Ilya Mazo$^3$ and Anton Yuryev$^3$,
Sergei Maslov$^{2}\footnote{To whom the correspondence should be addressed at
maslov@bnl.gov}$}
%\footnote{To whom the correspondence should be addressed:
%maslov@bnl.gov}
\affiliation{ $^1$ Department of Physics and
Astronomy, Stony Brook University, Stony Brook, New York, 11794, USA\\
$^2$Department of Condensed Matter Physics and Material Science,
Brookhaven National Laboratory, Upton, New York 11973, USA\\
$^3$Ariadne Genomics Inc., 9430 Key West Avenue, Suite 113,
Rockville, Maryland 20850, USA\\
%$^*$To whom the correspondence should be addressed at
%malov@bnl.gov
}
\date{\today}

\begin{abstract}
We develop a matrix-based approach to predict and verify indirect
interactions in gene and protein
%densely interconnected
regulatory networks.
It is based on the approximate transitivity of indirect regulations
(e.g. $A\rightarrow B$ and $B\rightarrow C$ often implies that $A\rightarrow C$)
and optimally takes into account the length of a
cascade and signs of intermediate interactions.
Our method is at its most powerful when applied to large and densely interconnected networks.
%,
%which are becoming prevalent due to rapid accumulation of data from
%high-throughput experiments.
It successfully predicts both the yet unknown
indirect regulations, as well as the sign (activation or repression)
of already known ones. The reliability of sign predictions
%of predicted
%as well as already known indirect interactions is
was calibrated using the
gold-standard sets of positive and negative
interactions.
We fine-tuned the parameters of our algorithm
by maximizing the area under the Receiver Operating Characteristic
(ROC) curve. We then applied the optimized
algorithm to large literature-derived networks of all direct and indirect regulatory
interactions in several model organisms
({\it Homo sapiens}, {\it Saccharomyces cerevisiae},
{\it Arabidopsis thaliana} and {\it Drosophila melanogaster}).
\end{abstract}

\maketitle

\noindent{{\large \bf Introduction}}

The development of high-throughput experimental
techniques lead to the accumulation of unprecedented amounts
of data describing regulatory interactions in model
organisms.  Effective computational algorithms are
needed to convert this treasure trove of information into
the system-wide understanding of the underlying biological
processes.

Regulatory interactions between proteins can be either direct or indirect. We would refer to a link from a regulatory protein to a target protein as {\it direct} if it is mediated by a direct molecular mechanism, such as e.g. transcriptional regulation of target protein's level by a transcription factor or phosphorylation of a substrate protein by a kinase. Conversely, regulations involving any number of intermediate proteins will be referred to as {\it indirect}. In fact, indirect regulations are vastly more common than the direct ones and thus are more likely to be detected experimentally. Large sets of regulatory interactions (both direct and indirect) are often represented
in terms of a directed network in which edges carry signs representing whether the regulation is an activation (positive sign) or an inhibition (negative sign). By ignoring the strength of interactions and combinatorial effects of several inputs such network provides a very simplified description of the real-life regulatory processes. 

In this work, we develop a novel algorithm which allows one to verify already known indirect regulations, infer their signs (if it is not known), and to predict the new ones, which have not yet been experimentally detected. As an input it uses a network consisting of all presently known 
regulatory interactions (both direct and indirect). Our algorithm also allows one to make an educated guess about which of the interactions in the original network are direct and which are indirect in cases when this information is not readily available (as e.g. in microarray experiments following a perturbation localized on one or several genes). Thus it contributes to a popular topic of
reconstructing direct regulatory network from
microarray data \cite{Friedman:JCB00,Peer:BI01}. Our algorithm works best when applied to large and heavily-interconnected networks. That is the reason we chose to apply it to networks in well-studied model organisms obtained using automatic text-mining technologies \cite{Novichkova:BI03}.

%@article{friedman2000ubn,
%  title={{Using Bayesian Networks to Analyze Expression Data}},
%  author={Friedman, N. and Linial, M. and Nachman, I. and Pe'er, D.},
%  journal={Journal of Computational Biology},
%  volume={7},
%  number={3-4},
%  pages={601--620},
%  year={2000}
%}

%@article{peer2001isp,
%  title={{Inferring subnetworks from perturbed expression profiles}},
%  author={Pe?er, D. and Regev, A. and Elidan, G. and Friedman, N.},
%  journal={Bioinformatics},
%  volume={17},
%  number={Suppl 1},
%  pages={S215--24},
%  year={2001},
%  publisher={Oxford Univ Press}
%}
%
%The sheer amount of information contained in these datasets (see the Table 1)
%and the necessity to take into account false positives
%caused e.g. by the variation of cell types and other
%experimental conditions used in different publications
%make most previous

Large-scale network analysis of indirect regulatory interactions
in yeast was recently studied in \cite{Wagner:BI01,Wagner:GB04,Kyoda:BI04}. These works focused on the classification of regulations as either direct or indirect and
subsequently pruning of indirect regulations. Pruning of indirect
regulations is a useful procedure from the point of network
simplification. However, being developed for relatively sparse
networks, these algorithms assume all links are equally reliable
and neither of these algorithms performs well for heavily
interconnected networks considered in this study.
%Due to the
%accruement of data from high-throughput experiments, heavily
%entangled networks are evidently unavoidable. Indeed, to
%effectively study large and heavily connected networks, one is
%forced to weigh links by their reliability. In principle, our
%algorithm for links verification could be efficiently used for
%pruning a network. In this manuscript, however, we choose to use
%it for prediction and verification of novel indirect regulations.\\
%%verification is used for predicitons
%

The emergent behavior of the rapidly growing body of knowledge
contained in regulatory and other biomolecular
networks was recently explored in a series of
publications of Rzhetsky and collaborators
\cite{Rzhetsky:NatBt05,Rzhetsky:PLoS106,Rzhetsky:PNAS06}.
%@article{cokol2005ebg,
%  title={{Emergent behavior of growing knowledge about molecular interactions}},
%  author={Cokol, M. and Iossifov, I. and Weinreb, C. and Rzhetsky, A.},
%  journal={Nat Biotechnol},
%  volume={23},
%  number={10},
%  pages={1243--7},
%  year={2005}
%}
%@article{rzhetsky2006mcc,
%  title={{Microparadigms: Chains of collective reasoning in publications about molecular interactions}},
%  author={Rzhetsky, A. and Iossifov, I. and Loh, J.M. and White, K.P.},
%  journal={Proceedings of the National Academy of Sciences},
%  volume={103},
%  number={13},
%  pages={4940--4945},
%  year={2006},
%  publisher={National Acad Sciences}
%}
%@article{rzhetsky2006scm,
%  title={{Self-correcting maps of molecular pathways.}},
%  author={Rzhetsky, A. and Zheng, T. and Weinreb, C.},
%  journal={PLoS ONE},
%  volume={1},
%  number={1},
%  pages={e61},
%  year={2006}
%}
The matrix-based approach advocated below nicely compliments
the Bayesian methods \cite{Rzhetsky:PLoS106}
of validation of large maps of biomolecular pathways
or, more generally, any set of published biological statements
\cite{Rzhetsky:PNAS06}.

The main idea behind our algorithm is as follows:
consider a protein $i$ regulating (either directly or indirectly)
a protein $k$ which in its turn is known to regulate (again
directly or indirectly) a protein $j$, then it is likely to
also have an indirect regulatory interaction between $i$ and $j$.
This simple observation could be further extended in two ways.
Firstly, indirect regulations could propagate along longer protein
cascades, thus a series of regulations $i\rightarrow k_1
\rightarrow k_2 \rightarrow j$ contributes to increase the likelihood of an
indirect regulation $i\rightarrow j$. Secondly, having multiple
parallel pathways reinforce the predictability. Therefore, if
a protein $i$ regulates proteins $k_1$, $k_2$ and each of them
regulates a protein $j$, it is even more likely to find an indirect
regulation from $i$ to $j$.

A simple-minded way to predict or verify an indirect regulation between
a protein $i$ and a protein $j$ is to simply count the number of
directed paths connecting $i$ and $j$. However, this counting scheme
does not take into account two important observations. First of all, paths
should be weighted differently according to their lengths. Inferences based on longer cascades is less reliable, and thus such should contribute less to the likelihood. We choose to exponentially discount longer paths
by weighting a path involving $n$ intermediate
proteins by a factor $\lambda^n$, where $\lambda<1$
is a parameter of our algorithm.
Secondly, the inferred sign of
the indirect regulation from different paths should agree with
each other. In general, if a protein $i$ and a protein $j$ are
connected by a multi-step path, the sign of the resultant
indirect regulation between $i$ and $j$ is given by the product of
signs of all intermediate edges. It is natural to assume that the
effect of a positive path (whose edges give a positive product)
and the effect of a negative path (whose edges give a negative
product) contradict and to some extent cancel each other.

In the next section, we will show that this central idea of
predicting likely indirect regulations could be easily
incorporated using a matrix formalism. Obviously, the likelihood
can serve as a quantitative measure of the reliability of any
regulation in a dataset. Thus one could also verify already known
regulations based on this calculated likelihood. A regulation with a high likelihood is deemed reliable. On the other hand,
indirect regulations  with a high likelihood
missing from the dataset could be reliably
predicted. %High-quality predictions are useful, and one would like
%to have as many as possible. However, we will show later in this
%paper,
As always, there is a tradeoff between the number of predictions
and their quality.
%Due to the usage of multi-step indirect paths
%from one protein to another, our method is tailored for heavily
%interconnected networks in which such multiple regulations are
%common.

We applied our algorithm to the set of genetic regulations
extracted from contents of
the entire PubMed database
(14,000,000 abstracts) and 47 full text journals. The automatic extraction
of interactions was made possible by the Medscan algorithm based on Natural Language Processing (NLP) techniques \cite{Novichkova:BI03,Daraselia:BI04}.
%@article{daraselia2004ehp,
%  title={{Extracting human protein interactions from MEDLINE using a full-sentence parser}},
%  author={Daraselia, N. and others},
%  journal={Bioinformatics},
%  volume={20},
%  number={5},
%  pages={604--611},
%  year={2004},
%  publisher={Oxford Univ Press}
%}
Both direct and
indirect regulatory interactions were collected for four model
organisms: {\it Homo sapiens}, {\it Saccharomyces cerevisiae},
{\it Arabidopsis thaliana} and {\it Drosophila melanogaster} (see
Table \ref{net_info} for details). As reflected in their
inter-connectedness index ($\mathrm{IC}=\langle k^{(in)}k^{(out)}\rangle/\langle k^{(in)}\rangle$), all these networks are globally
interconnected (IC$>1$). In particular, since the network of human
proteins is the largest and the most heavily interconnected (IC$\simeq$60)
among all networks used in this manuscript, we will show the results for this
network in more details.\\

\begin{table}
  \centering
  \caption{Regulatory networks in the four model organisms.
  The IC (inter-connectedness) index, defined
  as $\langle k^{(in)}k^{(out)}\rangle/\langle k^{(in)}\rangle$, measures how tightly
  bound together are the nodes in the  network. In this formula $k^{(in)}$ and $k^{(out)}$
  stand for the in- and out-degrees of nodes respectively. IC$>1$ means
  that the network is globally interconnected.
  From the table one can see that all of the networks used in this study
  are globally interconnected with the human dataset
  with IC$\simeq60$ being the most densely connected of them all.
  The gold-standard positive and negative sets consist of highly reproducible
  regulations (the ones reported in multiple publications)  with a given sign.\\}
  \label{net_info}
  \begin{tabular}{|c|c|c|c|c|c|c|}
  \hline
   & Number of & & \multicolumn{2}{c|}{Number of links}& \multicolumn{2}{c|}{Size of gold-standard set} \\
    \cline{4-7}
   Organisms & Proteins & \quad IC \quad \quad & positive & negative &  positive & negative \\
  \hline
  {\it Homo sapiens} & 7853 & 61.9 & 36426 & 16436 & 3442 & 1671\\
  {\it Saccharomyces cerevisiae} & 1218 & 3.42 & 1208 & 813 & 125 & 85\\
  {\it Arabidopsis thaliana} & 490 & 2.84 & 426 & 252 & 42 & 25 \\
  {\it Drosophila melanogaster} & 569  & 1.39 & 410 & 203 & 46 & 25\\
  \hline
  \end{tabular}
\end{table}

%Large-scale network analysis of indirect regulatory interactions
%in yeast was recently in (Wagner 2001, Tringe {\it et.al.} 2004
%and Kyoda {\it et.al.} 2004). These works focused on the
%classification of regulations as either direct or indirect and
%subsequently pruning of indirect regulations. Pruning of indirect
%regulations is a useful procedure from the point of network
%simplification. However, being developed for relatively sparse
%networks, these algorithms assume all links are equally reliable
%and neither of these algorithms performs well for heavily
%interconnected networks considered in this study. Due to the
%accruement of data from high-throughput experiments, heavily
%entangled networks are evidently unavoidable. Indeed, to
%effectively study large and heavily connected networks, one is
%forced to weigh links by their reliability. In principle, our
%algorithm for links verification could be efficiently used for
%pruning a network. In this manuscript, however, we choose to use
%it for prediction and verification of novel indirect regulations.
%%verification is used for predicitons

\noindent{{\bf \large Results and Discussion}}

\noindent{\bf Matrix formalism}

In this work, we represent the dataset of all known direct and indirect regulatory
interactions in a given organism as a directed network. In matrix
notation, it is fully defined by an adjacency matrix $A$ taking
the values
\begin{equation}
A_{ij}=\left\{ \begin{array}{l}
  +1 \quad \quad \mbox{if }i\mbox{ positively regulates }j,\\
  -1 \quad \quad \mbox{if }i\mbox{ negatively regulates }j,\\
  \;\;\;0  \quad \quad \mbox{if }i\mbox{ is not known to regulate }j.\\
\end{array}\right.
\end{equation}

To predict new indirect regulations and to quantify the
reliability of the existing ones, we use another matrix $X$ given by
\begin{eqnarray}
\label{series} 
X &=& A^2+\lambda A^3+\lambda^2 A^4+\lambda^3 A^5\cdots \nonumber\\
&=& \frac{A^2}{I-\lambda A}
\end{eqnarray}
where $\lambda<1$ is a parameter to be discussed later. $X_{ij}$
includes the contribution of all paths from $i$ to $j$.
$(A^n)_{ij}$ is the net number of paths (number of positive paths
minus the number of negative paths) of length $n$ from node $i$ to
node $j$, the sign of $X_{ij}$ is based on whether positive paths
or negative paths dominate. If positive (negative) paths dominate,
$X_{ij}$ is positive (negative), and it is likely that $i$ is
indirectly activating (repressing) $j$.

The constant $\lambda$ in Eq. (\ref{series}) is basically a free
parameter which could be optimized later to provide the best
performance for the algorithm. Generally speaking, $\lambda$
determines the weights of different paths. If $\lambda$ is chosen
to be less than one, the contribution from long paths is
exponential suppressed. In this work, we have chosen different
$\lambda$'s for different networks in order to optimize the
performance of our algorithm. We will first present our results
using the {\it optimal} value of $\lambda$. The definition of the
optimal $\lambda$ and its determination will be addressed later
on.\\

\noindent{\bf Calibration of reliability}

We have argued that the absolute magnitude of matrix elements of
$X$ is a measure of reliability of indirect regulations. Following
the matrix formalism, we calculate $X$ for four different
regulatory networks: {\it Homo sapiens}, {\it Saccharomyces
cerevisiae}, {\it Arabidopsis thaliana} and {\it Drosophila
melanogaster} (see the Materials and Methods section for additional
information).

In our algorithm, every non-zero element of $X$ possesses certain
predictive power. We collect all possible predictions by picking
out all non-zero $X_{ij}$'s. The validity of our algorithm is
evident if pairs $i$ and $j$ with large value of $|X_{ij}|$ are
likely to correspond to more reliable regulations. To show this is
indeed the case, one needs to use ``gold-standard set" containing
completely trustable regulations, which however is not readily
available. For this purpose, we define the gold-standard set to be
regulations which are frequently reported in the literature (for details of the cutoff on the number of publications, see Materials and Methods). The values of the median value of $|X|$ for all the non-zero matrix elements and those within the gold-standard
set are $3.9\times 10^{-3}$ and $3.5$ respectively.

%contrast to the median of $|X_{ij}|$comparing
%to $\langle |X_{ij}|\rangle_{\mbox{overall}}=0.116$, the average
%within the gold-standard set $\langle
%|X_{ij}|\rangle_{\mbox{golden}}=10.1$.

\begin{figure}
[htbp] \centering
\includegraphics*{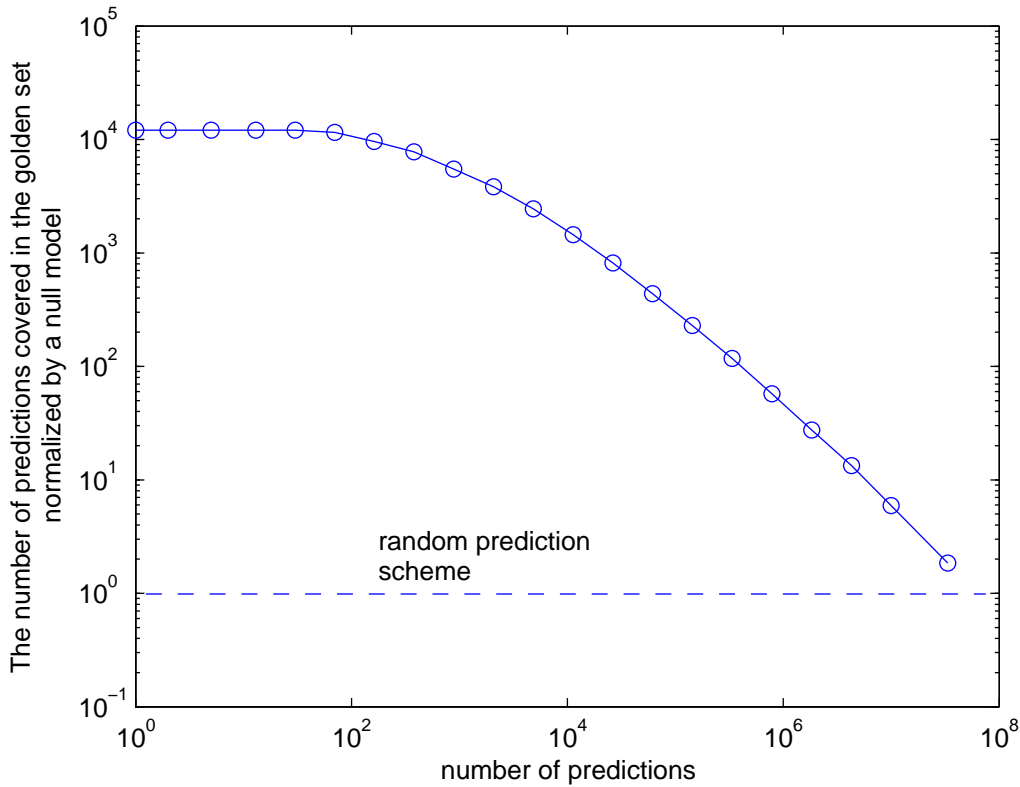}
\caption{The advantage of our prediction algorithm over null-model
expectations. The x-axis corresponds to the number $n$ of
predictions with the largest values of $|X_{ij}|$. The
y-axis is the ratio between the overlap of these $n$ predictions
with the combined (positive+negative) gold-standard
set and the null model expectation of this overlap.
One can see that our predictions are up to $10^4$
times more likely to  correspond to reliable,
experimentally verified regulations than expected by pure chance
alone.} \label{cf_rand}
\end{figure}

Figure \ref{cf_rand} shows a more detailed calibration of the matrix
elements. We define a predictive set of size $n$ using the $n$
predictions with the largest values of $|X_{ij}|$. If all the possible
predictions are used, the size of the set is huge (up to $10^7$).
The number of predictions covered in the gold-standard set is counted and
normalized by the corresponding number obtained by a set of $n$
random predictions. As shown in Figure \ref{cf_rand}, the overlap
between the gold-standard set and the best $100$ of our predictions is
$10,000$ (sic!) times better than what is expected by pure chance
alone. The advantage decreases when predictions with smaller
values of $|X_{ij}|$ are included. In case all possible
predictions are used, the predictive set is only sightly (2-fold) better
than a random set. This is expected since predictions with smaller
values of $|X_{ij}|$ are much less likely to be reliable.

Large $|X_{ij}|$ is a result of ``confirmation" by multi-step
paths from $i$ to $j$, therefore such predictions are likely to be
indirect in nature. To prove that it is indeed the case, we
separate the gold-standard set into direct and indirect subsets based on
the information obtained from literature as described. In agreement with our
expectation, the predictions are biased toward the indirect subset
(see Figure S1 in the Supporting Information).

\begin{figure}
[htbp] \centering
\includegraphics*{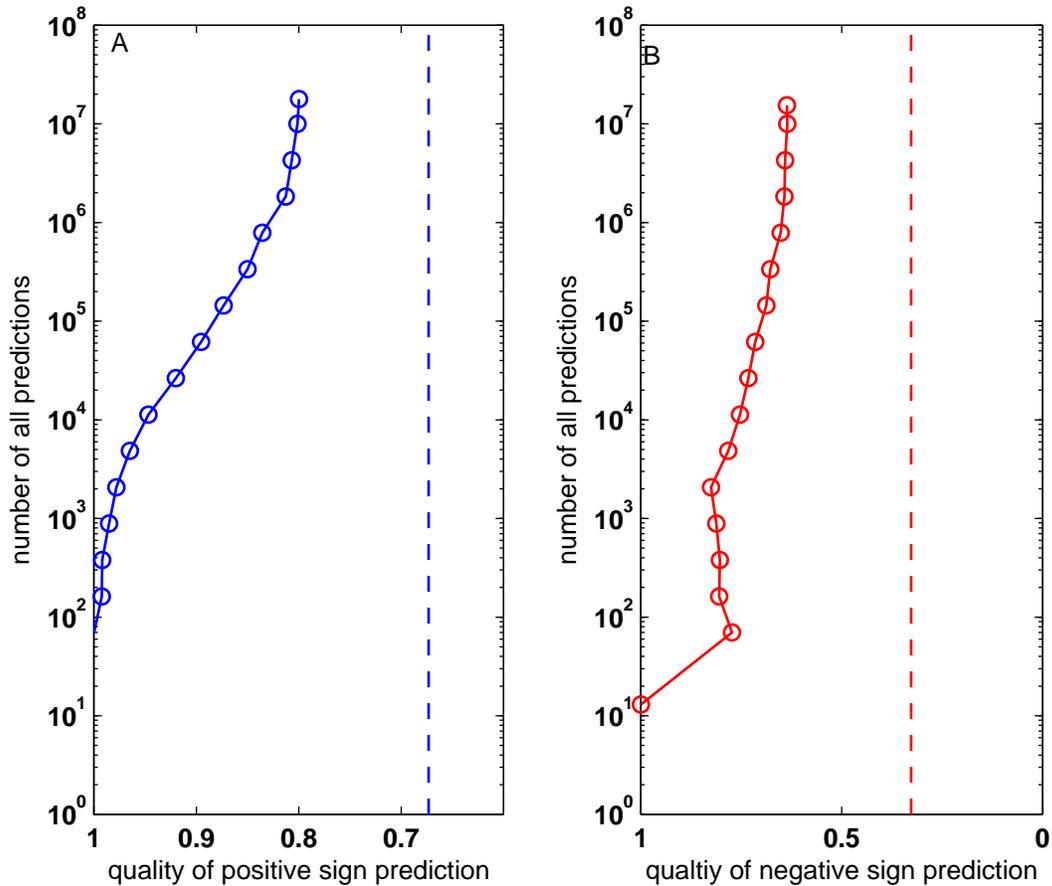}
\caption{The tradeoff between the number of predictions and their
average quality (panel A for positive predictions and B for
negative predictions). For a set of predictions, the average
quality is defined as the fraction of predictions whose sign
agrees with that in the gold-standard set. The dotted line is the quality
expected for a null model as described in the main text.}
\label{quality}
\end{figure}

Another use of matrix elements is to determine whether the
regulations are positive or negative. Under our formalism,
regulations corresponding to large {\it positive} matrix elements
are likely to represent positive regulations. In order to
calibrate the reliability for a set of predictions, we define the
average quality by counting the fraction of prediction whose
inferred sign agrees with that reported in the gold-standard set. Figure \ref{quality} shows the tradeoff between the number of predictions
and the average quality. As shown in Figure \ref{quality}A, a set of
predictions with average quality $100\%$ offers about $100$
predictions of positive regulation. However, if one is willing to
downgrade the quality to $95\%$, the number of predictions is up
to $5000$. By including all the positive entries in $X$, we are
offered a huge number of predictions, but with a relatively low
quality. However, even in that case, the average quality is still
much better than a null model, which is defined as the fraction of
positive regulations among all the regulations in the gold-standard set.
Thus the quality of our null model for positive (negative) regulations in human is
$3442/(3442+1671)=0.67$ ($1671/(3442+1671)$=0.33). They are shown as dashed lines in Figure \ref{quality}. Using negative
matrix elements, one could also predict negative regulations.
Large negative elements of $X$ are indeed more likely to have
negative signs in our gold-standard set (see Figure \ref{quality}B).

%%%%%%%%%%%%%

\begin{figure}[h!]
\centering
\includegraphics{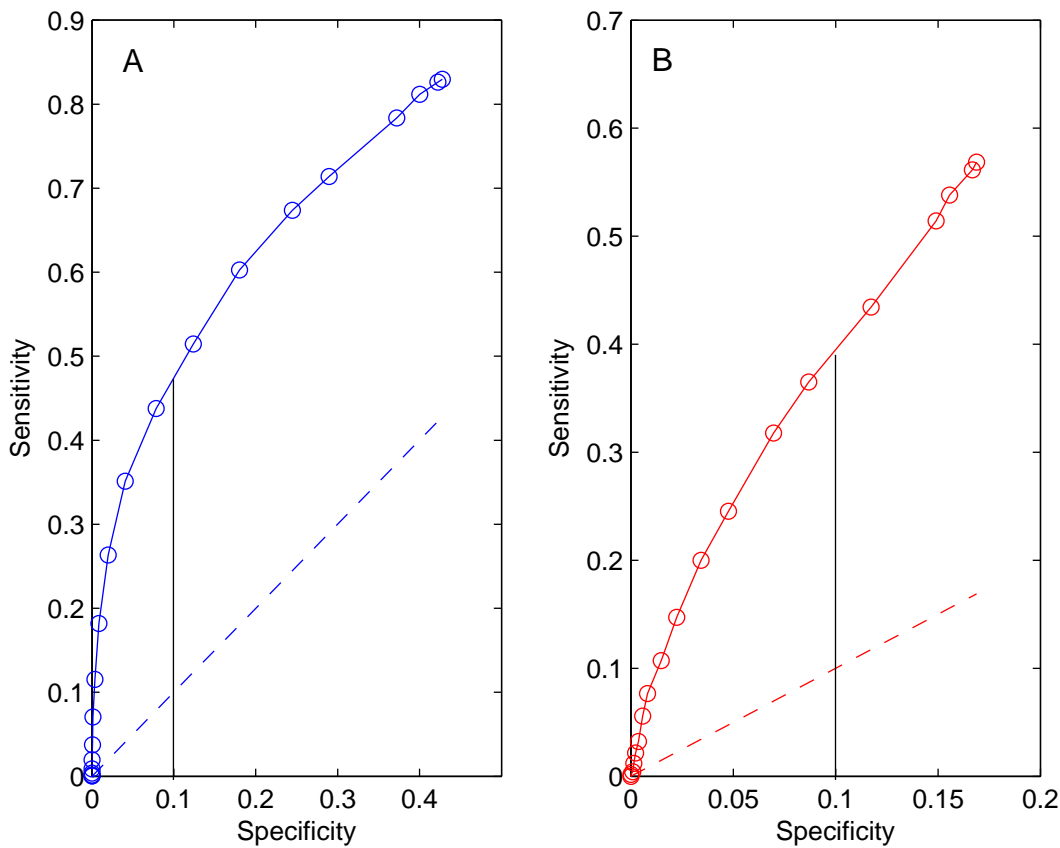}
\caption{ROC curves for sign predictions using positive $X_{ij}$ (panel A) and negative $X_{ij}$ (panel B). Each data point corresponds to a predictive set defined by a particular threshold of $X_{ij}$. The dotted lines are $y=x$, which is the null model expectations. The area under the ROC curve to the left of the solid line measures the
performance of our algorithm.} \label{ROC_1}
\end{figure}

To understand better the quality of our sign predictions, we study the Receiver Operating Characteristic (ROC) curves. Figure \ref{ROC_1}A is the ROC curve for
positive-sign predictions. It shows the sensitivity against specificity in different predictive sets as described by varying the $|X_{ij}|$ threshold. For positive-sign prediction, sensitivity is defined as the fraction of regulations in the positive gold-standard set which are predicted to be positive by our algorithm. Specificity, on the other hand, is defined as the fraction in the gold-standard {\it negative} set that are predicted to be positive by our algorithm. Data points close to the origin consist of predictions with large $X_{ij}$. The most important observation is the convexity of the curve, which means that the sign of interaction predicted by our method is more likely to be correct than expected by pure chance alone. In fact for a totally random predicted set, the ROC curve would be a straight line $y=x$. The area under a ROC curve is commonly used to quantify the performance of an algorithm. Using the negative $X_{ij}$ to predict negative regulations, one could similarly define sensitivity and specificity resulting another ROC curve as shown in Figure \ref{ROC_1}B.

Making use of the ROC curves, we could address the primary assumption
behind our definition of the gold-standard set: the larger is the number of
papers reporting a given interaction, the more reliable it is. We
define different gold-standard sets by varying the publication
cutoff. Gold-standard sets arising from a high cutoff are smaller in
size, but supposed to be more reliable. By comparing the area of
the ROC curves obtained from different gold-standard sets, we find that
indeed the ROC curve from a high-cutoff gold-standard set encloses a larger
area (see Figure S2 in Supporting information), which means those
regulations are indeed more trustable.\\

\noindent{\bf Validation of new predictions}

So far, every non-zero matrix element of $X$ stands for a
prediction. However, predictions could fall into two categories:
those covered in the gold-standard set and those not. Using the
predictions covered in the gold-standard set, we have calibrated the
reliability. Next, we are going to focus on the predictions
missing from the gold-standard set. First of all, we do not consider these
regulations as defects. In fact, being in the same predictive
set, they possess the same quality as those covered in the gold-standard
set. Therefore, we could use them as ``real" predictions of
missing regulations and expand the original dataset with these
predictions.

Table \ref{num_predictions} shows the number of the these new
predictions offered by our algorithm for the four model organisms.
Two different quality cutoffs $95\%$ and $75\%$ are used. The
number of predictions offered varies among the datasets, this is
because the datasets have different number of nodes, links and
topologies. However, in all cases, one could gain more predictions
by lowering the quality cutoff. We would like to stress that the
term ``quality" is calibrated separately in different datasets, therefore it
is not meaningful to compare the new predictions in human and
yeast even though their apparent qualities are the same. In fact,
predictions from human dataset are the most reliable, because our
algorithm is benefited from the heavily connected nature of the
human dataset.

Without experimental verification, it is hard to validate our new
predictions. To demonstrate our new predictions indeed make
biological sense, we compare our new predictions from human data to a complementary dataset of human regulatory interactions. The dataset is also obtained from literature using the Medscan algorithm but all the regulations
are not included in Table \ref{net_info} and the matrix $A$
(see the Materials and Methods section). We find that a significant fraction of
our new predictions coincide with this dataset. As shown in Table
\ref{num_predictions}, we have generated $2500$ new predictions
with an average quality of $95\%$ for the human network. Among them $750$ are indeed verified in the extra dataset. The corresponding P-value with respect to a random model is less than $10^{-100}$. The list of $2500$ predictions in human network, together with the predictions for other model organisms are listed in Table S3 in Supporting information.\\
%mean overlap: 1.4465, std=1.195

\begin{table}

  \centering
  \caption{Number of new predictions offered by our algorithm in
  regulatory networks of different organisms.\\}
  \label{num_predictions}
  \begin{tabular}{|c|c|c|}
  \hline
   Organisms  & 95\% sign quality & 75\% sign quality \\
  \hline
  {\it Homo sapiens}  & 2500 & $1.8\times 10^7$ \\
  {\it Saccharomyces cerevisiae}  & 190 & 7100\\
  {\it Arabidopsis thaliana}  & 85 & 13000 \\
  {\it Drosophila melanogaster}   & 650 & 1400\\
  \hline
  \end{tabular}
\end{table}

\noindent{\bf The optimal value of $\lambda$}

With ROC curves in hand, we are in a position to choose an
appropriate $\lambda$ for Eq. (\ref{series}). As a common practice, the quality of a ROC curve is quantified by the area under the curve.
The optimal $\lambda$ is thus the one whose ROC curve encloses the
largest area. However, the direct comparison of different areas
may be ambiguous. For example, compare the ROC curves from Fig.
\ref{ROC_1}, the one on the left panel encloses a larger area
while at the same time, the length covered in the x-axis is
longer. To overcome the problem, we introduce a cutoff in the
x-axis, and integrate area from $0$ up to the cutoff. In this
study, the cutoff is chosen to be $0.1$. As the beginning of the
ROC curve refers to the highly reliable predictions, the
introduction of the cutoff restricts ourselves in comparing the most 
reliable predictions. Thereafter, we define a quantity $\theta$ to
measure the overall performance of the algorithm, which is the
ratio between the area under the ROC curve from $0$ to the cutoff
and the corresponding area under the straight line $y=x$. The
ratio could be understood as the advantage of our algorithm over
random predictions.

\begin{figure}
[htbp] \centering
\includegraphics*{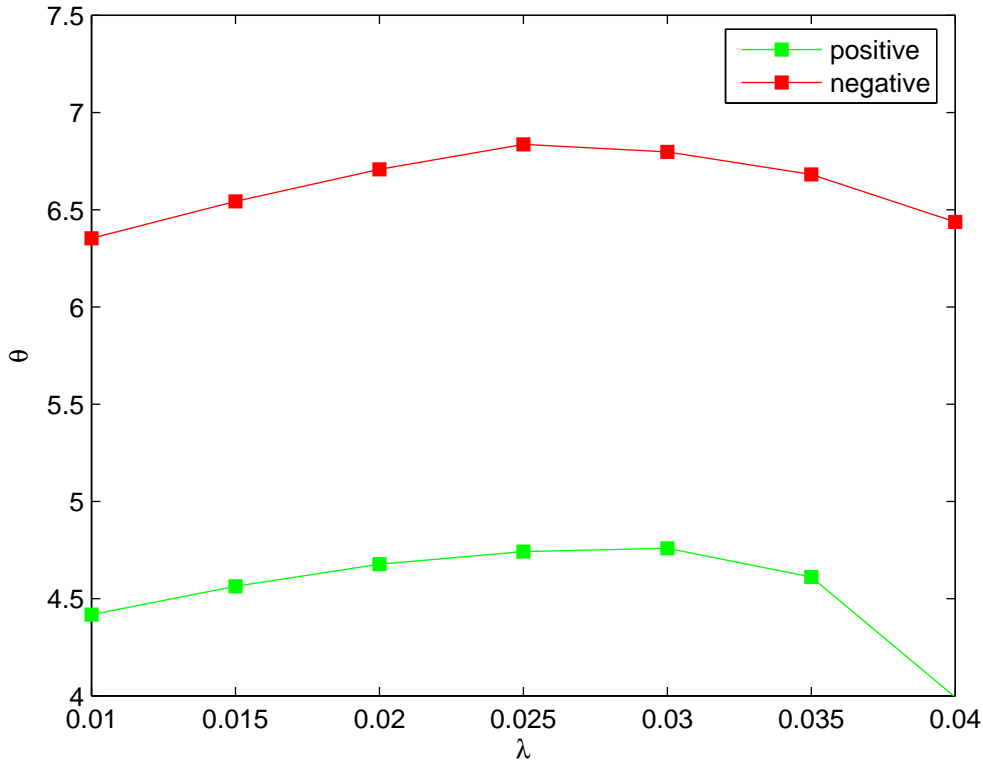}
\caption{Determination of the optimal value of $\lambda$. Optimal $\lambda$ maximizes $\theta$, defined by the ratio between the area under the
ROC curve from $0$ to $0.1$ and the corresponding area under the
straight line $y=x$. For human network, the optimal $\lambda$ for
positive and negative predictions are $0.025$ and $0.030$
respectively.} \label{theta_vs_lam}
\end{figure}

The performance of a particular $\lambda$ in Eq. \ref{series}
could be quantified by the resultant $\theta$. In Fig.
\ref{theta_vs_lam}, we plot $\theta$ against different $\lambda$'s
for positive and negative ROC curves in the human dataset. In
short, the optimal $\lambda$ is the one which gives the largest
$\theta$. From Figure \ref{theta_vs_lam}, the optimal $\lambda$ for
positive and negative predictions are $0.025$ and $0.030$
respectively. Readers are referred to the Materials and Methods section for
details of estimating $\theta$.\\
%do we have to write sth to discuss the choice of the cutoff? Supp?

\noindent{{\bf \large Materials and Methods}}

\noindent{\bf Collections of regulatory networks}

The regulatory networks for different model organisms are obtained
by the Medscan algorithm based on Natural Language Processing
(NLP). The term ``regulation" refers to the general influence of
the activity of one protein by another. Therefore, apart from
transcriptional regulations (which are direct regulations),
indirect regulations might be results of any cascades of post-transcriptional
or post-translational interactions between proteins.

Regulations are extracted from over 14 million PUBMED abstracts and 47 full text journals. Properties of regulations including the sign (positive or
negative) and its nature (direct and indirect) are parsed whenever
the information could be extracted from the corresponding
abstract. The number of times a regulation is reported in
literature is kept for the definition of gold-standard sets. Details of
each network is shown Table \ref{net_info}.

Apart from the data as shown in Table \ref{net_info}, we have
extracted an additional set ($35672$) of human regulations. The
regulations are not included with the datasets in Table
\ref{net_info} because their signs could not be parsed. In
this study, we use them as independent validation for the new
predictions generated by our algorithm.\\

\noindent{\bf Definition of gold-standard sets}

For each organism, the corresponding positive (negative) gold-standard set
is defined by the top $10\%$ most frequently reported positive
(negative) regulations. The size of each gold-standard set could be found
in Table \ref{net_info}. For human dataset, the publication cutoffs used in positive and negative gold-standards are $8$ and $5$ respectively.\\
%Comparing the whole human dataset
%and its corresponding gold-standard set, the average number of times that
%an interaction is reported are $3.35$ and $22.6$ respectively. The
%ratios between the two numbers are roughly the same for the
%other organisms. GIVE THE CUTOFFS\\
%may be sth supp information...

\noindent{\bf Estimation of the area under a ROC curve}

For each ROC curve, we fit the data point by the function $y=Ax^B$
using the MATLAB function \verb"fminsearch", which is based on the
Nelder-Mead method in non-linear optimization. The area under the
fitted curve is numerically evaluated in MATLAB by the function
\verb"quadl" using the adaptive Lobatto quadrature.

To exclude the data points far from the origin, which are results
of less reliable predictions, we introduce a cutoff in the x-axis.
Area is integrated from $0$ up to the cutoff. In this study, a
cutoff of value $0.1$ is used.\\

\noindent{\bf Acknowledgements}

Work at Brookhaven National Laboratory was carried out under Contract No. DE-AC02-98CH10886, Division of Material Science, U.S. Department of Energy. This work was supported by National Institute of General Medical Sciences Grant 1 R01 GM068954-01. IM thanks the Institute for Strongly Correlated and Complex Systems at Brookhaven National Laboratory for hospitality and financial support during visits when the majority of this work was done.  KKY and SM visit to Kavli Institute for Theoretical Physics where part of this work was accomplished was supported by the National Science Foundation under Grant No. PHY05-51164.

%\noindent{\large \bf Acknowledgements}

%\noindent{\large \bf References}
\bibliographystyle{plos}
\bibliography{yan_indirect_regulation_Nov07}

\begin{thebibliography}{10}
\providecommand{\url}[1]{\texttt{#1}}
\providecommand{\urlprefix}{URL }
\expandafter\ifx\csname urlstyle\endcsname\relax
  \providecommand{\doi}[1]{doi:\discretionary{}{}{}#1}\else
  \providecommand{\doi}{doi:\discretionary{}{}{}\begingroup
  \urlstyle{rm}\Url}\fi
\providecommand{\bibAnnoteFile}[1]{%
  \IfFileExists{#1}{\begin{quotation}\noindent\textsc{Key:} #1\\
  \textsc{Annotation:}\ \input{#1}\end{quotation}}{}}
\providecommand{\bibAnnote}[2]{%
  \begin{quotation}\noindent\textsc{Key:} #1\\
  \textsc{Annotation:}\ #2\end{quotation}}
\providecommand{\eprint}[2][]{\url{#2}}

\bibitem{Friedman:JCB00}
Friedman N, Linial M, Nachman I, Pe'er D (2000) Using bayesian networks to
  analyze expression data.
\newblock J Comput Biol 7:601--620.
\newblock \doi{10.1089/106652700750050961}.
\bibAnnoteFile{Friedman:JCB00}

\bibitem{Peer:BI01}
Pe'er D, Regev A, Elidan G, Friedman N (2001) Inferring subnetworks from
  perturbed expression profiles.
\newblock Bioinformatics 17 Suppl 1:S215--24.
\bibAnnoteFile{Peer:BI01}

\bibitem{Novichkova:BI03}
Novichkova S, Egorov S, Daraselia N (2003) Medscan, a natural language
  processing engine for medline abstracts.
\newblock Bioinformatics 19:1699--1706.
\bibAnnoteFile{Novichkova:BI03}

\bibitem{Wagner:BI01}
Wagner A (2001) How to reconstruct a large genetic network from n gene
  perturbations in fewer than n(2) easy steps.
\newblock Bioinformatics 17:1183--1197.
\bibAnnoteFile{Wagner:BI01}

\bibitem{Wagner:GB04}
Tringe SG, Wagner A, Ruby SW (2004) Enriching for direct regulatory targets in
  perturbed gene-expression profiles.
\newblock Genome Biol 5:R29.
\newblock \doi{10.1186/gb-2004-5-4-r29}.
\bibAnnoteFile{Wagner:GB04}

\bibitem{Kyoda:BI04}
Kyoda K, Baba K, Onami S, Kitano H (2004) Dbrf-megn method: an algorithm for
  deducing minimum equivalent gene networks from large-scale gene expression
  profiles of gene deletion mutants.
\newblock Bioinformatics 20:2662--2675.
\newblock \doi{10.1093/bioinformatics/bth306}.
\bibAnnoteFile{Kyoda:BI04}

\bibitem{Rzhetsky:NatBt05}
Cokol M, Iossifov I, Weinreb C, Rzhetsky A (2005) Emergent behavior of growing
  knowledge about molecular interactions.
\newblock Nat Biotechnol 23:1243--1247.
\newblock \doi{10.1038/nbt1005-1243}.
\bibAnnoteFile{Rzhetsky:NatBt05}

\bibitem{Rzhetsky:PLoS106}
Rzhetsky A, Zheng T, Weinreb C (2006) Self-correcting maps of molecular
  pathways.
\newblock PLoS ONE 1:e61.
\newblock \doi{10.1371/journal.pone.0000061}.
\bibAnnoteFile{Rzhetsky:PLoS106}

\bibitem{Rzhetsky:PNAS06}
Rzhetsky A, Iossifov I, Loh JM, White KP (2006) Microparadigms: chains of
  collective reasoning in publications about molecular interactions.
\newblock Proc Natl Acad Sci U S A 103:4940--4945.
\newblock \doi{10.1073/pnas.0600591103}.
\bibAnnoteFile{Rzhetsky:PNAS06}

\bibitem{Daraselia:BI04}
Daraselia N, Yuryev A, Egorov S, Novichkova S, Nikitin A, et~al. (2004)
  Extracting human protein interactions from medline using a full-sentence
  parser.
\newblock Bioinformatics 20:604--611.
\newblock \doi{10.1093/bioinformatics/btg452}.
\bibAnnoteFile{Daraselia:BI04}

\end{thebibliography}

\newpage
\noindent{\bf \large Supporting Information}

\noindent{Figure S1. The coverage of direct and indirect gold-standard sets. The coverage of the direct (indirect) subset for a given set of predictions is defined as the number of verified predictions normalized by the size of the direct (indirect) gold-standard set. Data points closer to the origin refer to predictions with larger average value of $|X_{ij}|$. As reflected by the convexity of the curve, those regulations are more likely to be indirect rather than direct.}\\

\noindent{Figure S2. ROC curves of the human regulatory network using gold-standard
sets with different cutoffs. A gold-standard set is defined by regulations
which are highly reported in literature. An interaction belonging to the gold-standard set with cutoff $5\%$ is among the top $5\%$ of the dataset in terms of the number of papers reporting. Data points labeled by $\circ$, $\triangle$ and $\star$ are the results of gold-standard sets whose sizes are $5\%$, $10\%$ and $20\%$ of the original network. These correspond to publication cutoffs $14, 8, 4$ for positive regulations and $9, 5, 3$ for negative regulations respecitively. The ROC curves (positive and negative) corresponding to a high-cutoff gold-standard set enclose larger areas.}\\

\noindent{Table S3. The new predictions and their signs offered by our algorithm with an average quality of 95\%. (\texttt{Table\_S3.xls})}

\begin{figure}[h!]
\centering
\includegraphics{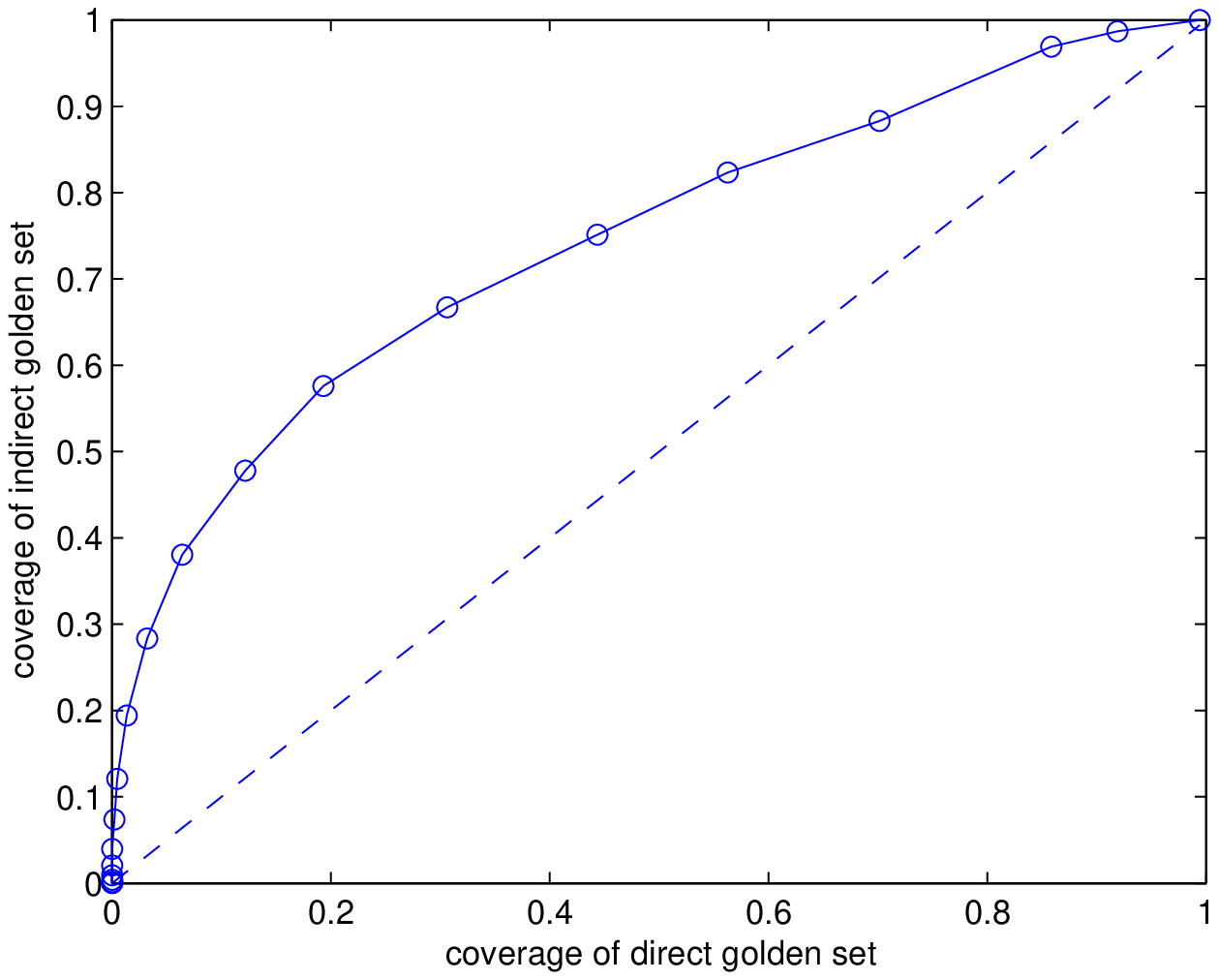}
\end{figure}

\begin{figure}[h!]
\centering
\includegraphics{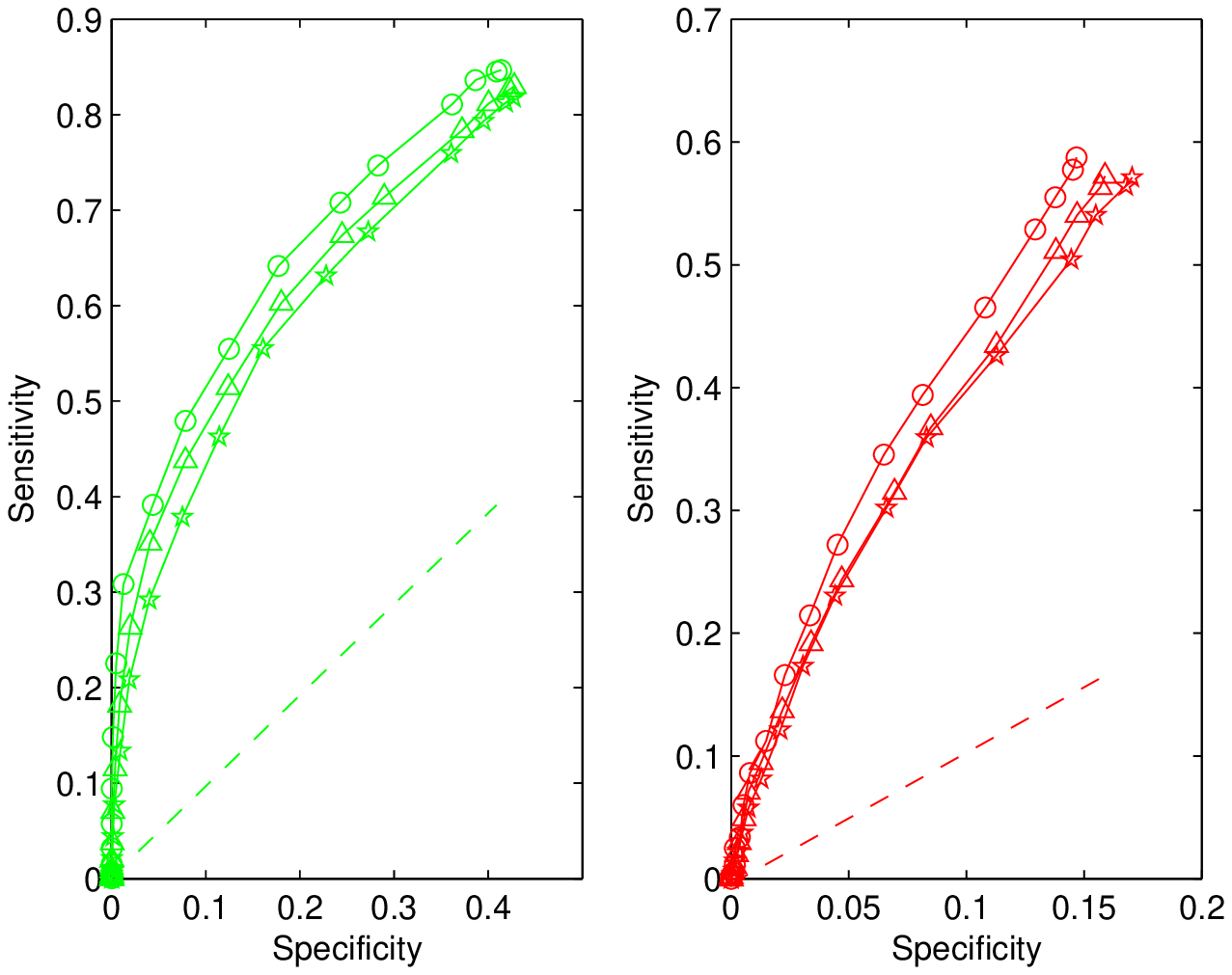}
\end{figure}

\end{document}